\documentclass{an}
\usepackage{times}
\usepackage{graphicx}

\begin{document}

\title{Photometric study of two $\beta$~Cephei pulsators in eclipsing systems}

\author{
    D.~Drobek\inst{1} \and
    A.~Pigulski\inst{1} \and
    R.~R.~Shobbrook\inst{2} \and
    A.~Narwid\inst{1}
}

\institute{
    Instytut Astronomiczny Uniwersytetu Wroc{\l}awskiego, Wroc{\l}aw, Poland \and
    Research School of Astronomy and Astrophysics, Australian National University, Canberra, ACT, Australia
}

\keywords{stars: oscillations -- binaries: eclipsing -- stars: individual (HD\,101794, HD\,101838)
-- open clusters and associations: individual (Stock 14)}

\abstract{%
We present results of a~photometric study of the young southern open cluster Stock~14. This
cluster is known to contain two eclipsing systems with presumed $\beta$~Cephei components,
HD\,101794 and HD\,101838. We confirm variability due to pulsations and eclipses in both targets
and announce the discovery of other variable stars in the observed field.}
\maketitle

\section{Introduction}
Pulsating components of
eclipsing binary systems play an
important role in asteroseismology. This is because a combination of photometric
and spectroscopic solutions allows determination
of masses and radii of both components. These parameters can be subsequently
used in seismic
study. By putting tighter constraints on fundamental stellar parameters,
greater accuracy of the asteroseismic mo\-del can be achieved.

Photometric variability of HD\,101794 was discovered by the Hipparcos satellite. The
star was at first identified as a $\gamma$~Cassiopeiae-type variable and was assigned the
name V916 Cen. Its spectral type was classified as B1.5~V (Feast et al.~1961),
B0/1\,ne (Houk \& Cowley 1975), B1\,IVne (Garrison, Hiltner \& Schild 1977) and
B0.5\,IVne (FitzGerald \& Miller 1983). Emission was confirmed by measurements of $\beta$
index (Moffat \& Vogt 1975; Klare \& Neckel 1977; Johannson 1981; Kaltcheva 2003)
and H$\alpha$ photometry (McSwain \& Gies 2005). The eclipsing nature of HD\,101794
was revealed by Pojma{\'n}ski (2000), who determined its orbital period
$P_{\mathrm{orb}}=$ 1.4632~d from the $I$-filter All Sky Automated Survey phase 2 (ASAS-2) photometry.
Since both eclipses are similar in depth, the system could be a~double-lined (SB2) spectroscopic
binary. Ana\-ly\-sis of the ASAS-3 (Pojma{\'n}ski 2002) $V$-filter photometry by
Pigulski \& Poj\-ma{\'n}ski (2008) led to the discovery of two pulsation modes:
$f_1=$ 4.45494~d$^{-1}$, typical for a~$\beta$~Cephei star, and
$f_2=$ 1.83952~d$^{-1}$, attributed to $\lambda$~Eridani-type variability by
the authors. It is not clear if both modes ori\-gi\-nate in the same component.

HD\,101838 is also an early B-type star, classified as B1\,III (Feast et al.~1961),
B0.5/1\,III (Houk \& Cowley 1975), B1\,II-III (Garrison et al.~1977) and
B0\,III (FitzGerald \& Mil\-ler 1983). It was discovered to be variable by
Pigulski \& Poj\-ma{\'n}ski (2008). The star exhibits a single pulsation mode with
frequency $f_1=$ 3.12764~d$^{-1}$, which is quite low for a~$\beta$~Cephei
variable. The orbital period probably amounts to 5.41167~d, but from
the photometry alone one may not exclude the possibility that the true orbital period is
twice as long. If the shorter period is assumed, the eclipsing light curve does not
exhibit a~secondary minimum. If the period is longer, both eclipses are similar
in depth. The latter case would be more favourable, as the system would probably
be an~SB2 binary. Additional observations, preferably spectroscopic, are needed
to resolve this ambiguity.

It is worth noting that both HD\,101794 and HD\,101838 might be members
of Stock~14, a~sparsely populated young open cluster
in Centaurus. According to the WEBDA\footnote{Web version of the BDA (Base de Donn{\'e}es des
Amas ouverts) database, http://www.univie.ac.at/webda} da\-ta\-ba\-se, the mean
$E(B-V)$ colour excess for cluster stars amounts to 0.23~mag, the distance of the
cluster is about 2.1~kpc and it is 11\,$\times$\,10$^6$
years old. Arguments in favour of the membership of HD\,101794 and HD\,101838 can be found in papers
by Moffat \& Vogt (1975), FitzGerald \& Miller (1983) and Peterson \& FitzGerald (1988).
The membership criteria used in those papers are either based on photometry (position in colour-magnitude
and colour-colour diagrams, the agreement of individual reddenings), or on the concordance of
spectral types of individual stars.

Due to the known poor spatial resolution of the ASAS observations, amounting to about
15 arcseconds per pixel, it is higly desirable to carry out follow-up
photometry in order to minimize the possibility of contamination by nearby stars.
The aim of this paper is to verify that the variability
of HD\,101794 and HD\,101838 is both eclipsing and pulsating in nature, and also to
provide a~status report on the ongoing photometric study of the Stock~14 field.

\section{Observations, calibration and reduction}
Follow-up observations were carried out between 7~March and 13~May 2007 (19 nights in
total). We used the 40-inch Cassegrain telescope at Siding Spring Observatory, Australia.
The detector was the Wide Field Imager (WFI), an eight CCD (2048\,$\times$\,4096
pixels each) mosaic with 52\,$\times$\,52 arcmin$^2$ field of view. The entire target cluster
is located in a single WFI chip \#3. For this reason, only results from that particular CCD
are presented here. A~set of standard $UBV$ Johnson filters was used, with respectively
881, 1085, and 1125 frames taken through each filter. Since target stars are bright,
typical exposure times were short: starting from 10~s in $V$ up to 50~s in $U$. In total, we
identified 7011 stars in our $V$-filter reference frame.

The frames were calibrated using a~standard procedure, which included correcting signal
values in each CCD pixel for the known
non\-li\-nea\-ri\-ties\footnote{Tinney, C.G.: 2002, http://www.aao.gov.au/wfi/wfi\_pfu.html}.
Stellar magnitudes
were calculated using the Daophot software package (Stetson 1987). At\-mo\-sphe\-ric effects
were accounted for by means of differential photometry using 15 comparison stars,
distributed evenly in the frame. The mean instrumental magnitude of each star was calculated as
an average value from its $\sigma$-clipped time-series. In eclipsing binaries, only points
outside of eclipses were considered in this computation. Judging by the errors in mean
instrumental magnitudes and by the appearance of colour-magnitude and co\-lour-co\-lour
diagrams we conclude that a~subset of 2401 stars has reasonably good photometry in both
$V$~and $B$~filters. Only 943 stars have decent photometry in all filters. We
have not obtained any photometry of star V810~Cen. This supergiant is the brightest star
in the field and was severely overexposed in all frames.
Photometric standards were not observed during our observing run. However, we have been
able to transform our instrumental magnitudes to the standard system using data from
the paper by Peterson \&~FitzGerald (1988).

\section{Analysis and results}
The time series of HD\,101794 and HD\,101838 were ana\-lysed in the following way. First,
they were folded with orbital periods provided by Pigulski \& Pojma{\'n}ski (2008), i.e.,
1.46323~d and 5.41167~d, respectively. Visual inspection of the resulting light curves
revealed the eclipses in both targets. Significant scatter of data points was
also present. That was to be expected, taking into account the existence of pulsations.
In order to proceed with the analysis we had to remove the contribution of eclipses.
The orbital frequency with the appropriate number of harmonics was fitted to the original
data. Finally, a~discrete Fourier transform of the residuals was
calculated. This step revealed the presence of pulsations.

\begin{figure}[!ht]
\includegraphics[width=3.2in]{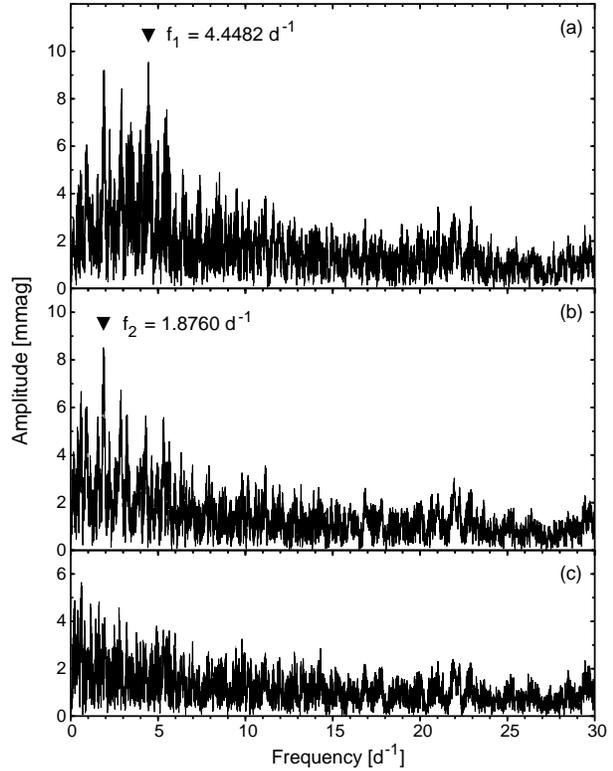}
\caption{Frequency spectrum of the $V$-filter photometric data of HD\,101794:
(a) after subtraction of the eclipsing light curve;
(b) prewhitened with $f_1$; (c) prewhitened with $f_1$ and $f_2$.}
\label{hd794trf}
\end{figure}

\begin{figure}[!ht]
\includegraphics[width=3.2in]{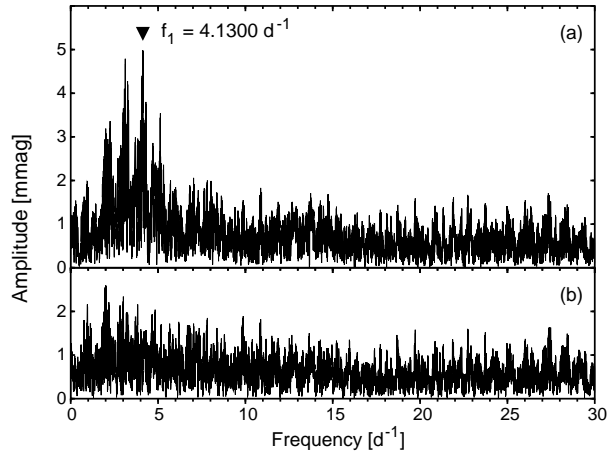}
\caption{Frequency spectrum of the $V$-filter photometric data HD\,101838: (a) after subtraction
of the eclipsing light curve; (b) prewhitened with $f_1$.}
\label{hd838trf}
\end{figure}

The most prominent maximum visible in the frequency spectrum of HD\,101794 corresponds
to $f_1=$ 4.4482~d$^{-1}$ (Fig.~\ref{hd794trf}), which is in good agreement with $f_1$ of Pigulski
\&~Pojma{\'n}ski (2008). The second frequency we detect, $f_2^\prime=$ 1.8760~d$^{-1}$ is a 25-day alias
of $f_2$ from aforementioned paper (our photometry consisted of three runs separated rou\-gh\-ly
by 25--30~d). We do not detect any additional
frequencies in the residuals.

The frequency spectrum of HD\,101838 (Fig.~\ref{hd838trf}) features two near\-ly equal maxima. The one at
$f=$ 3.1299~d$^{-1}$ corresponds to the frequency already known from the ASAS-3 data,
the other is its daily alias. The amplitude of the latter is slightly larger than the former.
With no means to decide which maximum corresponds to the real frequency, we assumed that
$f_1=$ 4.1300~d$^{-1}$ is the true one. After pre\-white\-ning the spectrum with
this frequency, no significant maxima can be seen.

\begin{figure}
\includegraphics[width=3.2in]{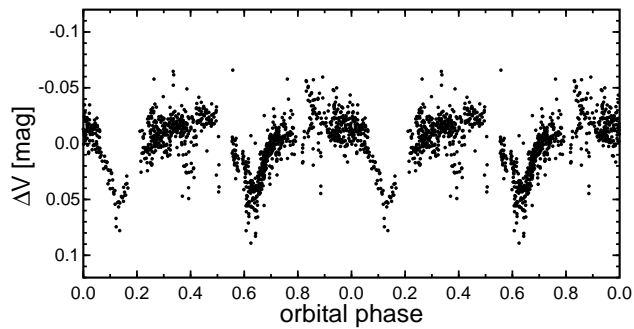}
\caption{Eclipsing differential $V$-filter light curve of HD\,101794. The orbital period is 1.46323~d.}
\label{hd794ecl}
\end{figure}

\begin{figure}
\includegraphics[width=3.2in]{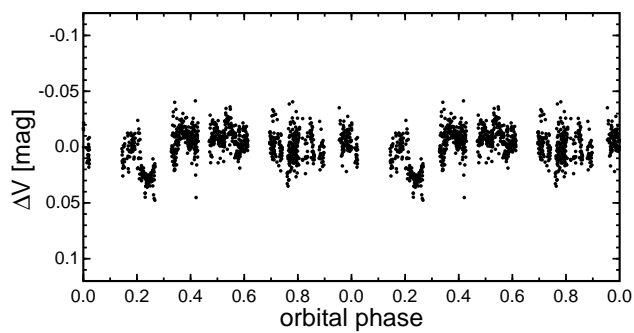}
\caption{Eclipsing differential $V$-filter light curve of HD\,101838. The orbital period is 5.41167~d.
A~shallow primary eclipse is seen around the orbital phase of 0.25.}
\label{hd838ecl}
\end{figure}

Having removed the contribution of pulsations from our original time series, we folded them
with their respective ASAS-3 orbital periods. Resulting eclipsing light curves of HD\,101794
and HD\,101838 are shown in Fig.~\ref{hd794ecl} and Fig.~\ref{hd838ecl}, respectively.
There is no doubt that the targets indeed
are eclipsing binary systems. While HD\,101794 shows two ec\-lip\-ses of similar depth, only a~shallow
primary eclipse is seen in HD\,101838.

An interesting by-product of this work is the discovery of seven new variable stars in the field
of the Stock 14 open cluster. They include a multiperiodic $\beta$~Cephei star, a multiperiodic SPB star
and a~$\delta$~Scuti variable with at least two excited modes. Since only frames from one CCD
chip of the WFI have been analysed so far, the variability search is not complete. Because of this,
we refrain from disclosing detailed information about discovered variable stars in this paper.
The full results of the variability search in the field of Stock 14 will be published elsewhere.

\acknowledgements
This research has made use of the WEBDA database, operated at the
Institute for Astronomy of the University of Vienna, the VizieR catalogue access tool, CDS,
Strasbourg, France and the SAOImage DS9, developed by Smithsonian
Astrophysical Observatory. The authors are grateful to the EC for the establishment of the
European Helio- and Asteroseismology Network HELAS, which made their participation at this conference
possible.  The work was supported by the N\,N203\,302635 grant from MNiSzW.

\end{document}